\documentclass[preprint,preprintnumbers,amsmath,amssymb,superscriptaddress]{revtex4}

\usepackage{bm}
\usepackage{graphicx}
\usepackage{subfig}
\usepackage{array}
\usepackage{color}
\usepackage[section]{placeins}
\allowdisplaybreaks[3]

\graphicspath{{figures/}}

\makeatletter

\newcommand{\Rmnum}[1]{\expandafter\@slowromancap\romannumeral #1@}
\makeatother

\begin{document}

\title{$B \rightarrow TT$ decays in the QCD factorization approach}

\author{Ya-Bing Zuo}\email{zuoyabing@lnnu.edu.cn}\quad
\author{Jia-Yu Zou} \quad
\author{Shi-Yu Liang} \quad
\author{Ming-Ge Li} \quad
\author{Shan-Shan Hu}
\affiliation{Department of physics, Liaoning
Normal University, Dalian 116029, P.R.China }
\affiliation{Center for Theoretical and Experimental High Energy Physics, Liaoning Normal University, Dalian 116029, P.R.China}

\begin{abstract}

In this study, the nonleptonic two-body $B$ decays into two tensor mesons (including $a_2(1320)$, $K^*_2(1430)$, $f_2(1270)$, $f^\prime_2(1525)$, denoted generically as $T$) are investigated in the QCD factorization approach. The branching ratios, longitudinal polarization fractions, and CP asymmetries are predicted systematically. It is found that, from the perspective of central values, the branching ratios for the decays $B \rightarrow \{a_2 f^\prime_2, f_2 f^\prime_2\}$, $\{a_2f_2, f_2 f_2\}$ and $\{f_2 K^*_2, K^*_2f^\prime_2\}$ are at the order of $10^{-8}$, $10^{-7}$ and $10^{-6}$, respectively. The branching ratios for the decays into $\{a_2 a_2, K^*_2 K^*_2\}$ and $a_2 K^*_2$ are at the order of $10^{-8} - 10^{-7}$ and $10^{-7} - 10^{-6}$, respectively. The longitudinal polarization fractions are approximately $0.7-1$. Especially, for the $\bar{B}^0/B^0 \rightarrow \{K^{*-}_2 K^{*+}_2, f^\prime_2 f^\prime_2\}$ decays, the longitudinal polarization fractions are equal to $1$. The CP asymmetry for the $B^\pm \rightarrow a^\pm_2 f_2$ modes is most significant, roughly $-22\%$. The CP asymmetries for the decays $B \rightarrow \{a_2 f^\prime_2, f_2 f^\prime_2, f^\prime_2 f^\prime_2\}$ and $\bar{B}^0/B^0 \rightarrow K^{*-}_2 K^{*+}_2$ are equal to zero. Our results may be tested by more precise experiments in the future.

\end{abstract}

\maketitle

\section{Introduction}

The nonleptonic two-body $B$ meson decays have been studied extensively in the past two decades both theoretically and experimentally. Since many charmless hadronic $B$ decays involving a light tensor meson in the final states have been measured in experiments\cite{PDG2024}, tensor mesons get more attentions. In the quark model, the low-lying tensor meson with $J^{PC}=2^{++}$ can be modeled as a constituent quark-antiquark pair with total orbital angular momentum $L=1$ and spin $S=1$. The observed isovector mesons $a_2(1320)$, isodoublet states $K^*_2(1430)$ and two isosinglet mesons $f_2(1270)$, $f^\prime_2(1525)$ form an light flavor $SU(3)$ $1^3 P_2$ nonet. Compared with the pseudoscalar and vector mesons, the tensor mesons are more complicated. Up to now, the $B \rightarrow TP, TV$ ($P$, $V$ and $T$ denote the light pseudoscalar, vector and tensor mesons, respectively.) decays have been investigated via several approaches, such as the naive factorization\cite{NaiveF1,NaiveF2,NaiveF3,NaiveF4,NaiveF5,NaiveF6}, QCD factorization\cite{QCDF} and perturbative QCD\cite{pQCD1,pQCD2,pQCD3}.
Recently, the $B_s \rightarrow TT$ decays were also analyzed in the perturbative QCD approach\cite{pQCDBs}.

Theoretical calculations of exclusive decays require the knowledge of nonperturbative QCD, which are generally parameterized in terms of decay constants, form factors and some nonfactorizable contributions. For heavy hadrons containing one heavy quark, it is useful to adopt the heavy quark effective field theory (HQEFT), in which a heavy quark expansion is performed and calculation of nonperturbative QCD effects can be greatly simplified\cite{HQEFT1,HQEFT2,HQEFT3}. The $B_{(s)} \rightarrow T$ form factors were calculated via the light cone sum rules (LCSR) in HQEFT and used to investigate the corresponding semileptonic decays\cite{Bto2ppSL}.
In this study, we intend to investigate the nonleptonic $B$ decays into two tensor mesons by using the QCD factorization approach with these form factors as inputs, predicting the branching ratios, longitudinal polarization fractions and CP asymmetries.

The remaining part of this paper is organized as follows. In Sec.II, we give the physical properties of tensor mesons used in our calculations such as the decay constants, form factors and light cone distribution amplitudes (LCDAs).  Then we work out the next to leading order corrections to $B \rightarrow TT$ decays in Sec.III, and present numerical results and discussions in Sec.IV.
Section V is our summary.

\section{Physical properties of tensor mesons}\label{SecII}

\subsection{Tensor mesons}

The observed $J^{PC}=2^{++}$ tensor mesons $a_2(1320)$, $K^*_2(1430)$, $f_2(1270)$, $f^\prime_2(1525)$ form a light flavor $SU(3)$ $1^3 P_2$ nonet.
The flavor contents for isovector and isodoublet tensor resonances are obvious.
Just like the $\eta-\eta^\prime$ mixing in the pseudoscalar case, the isosinglet tensor states $f_2(1270)$ and $f^\prime_2(1525)$ also have a mixing, and their wave functions are defined as
\begin{eqnarray}
& & f_2(1270) = \frac{1}{\sqrt{2}} ( f^u_2 + f^d_2) \cos \theta_{f_2} + f^s_2 \sin \theta_{f_2}, \\
& & f^\prime_2(1525) = \frac{1}{\sqrt{2}} ( f^u_2 + f^d_2) \sin \theta_{f_2} - f^s_2 \cos \theta_{f_2},
\end{eqnarray}
with $f^u_2 = u \bar{u}$ and likewise for $f^{d,s}_2$. Since $\pi \pi$ is the dominant decay mode of $f_2(1270)$
whereas $f^\prime_2$ decays predonimantly into $K\bar{K}$\cite{PDG2024}, it is obvious that this mixing angle should be small. More precisely, it is found that $\theta_{f_2} = 7.8^\circ$\cite{f2mix}. Therefore, $f_2(1270)$ is primarily a $(u\bar{u} + d\bar{d})/\sqrt{2}$ state, while $f^\prime_2(1525)$ is dominantly $s\bar{s}$. For simplicity, we shall neglect this mixture in our calculations.

For a tensor meson, the polarization tensors $\epsilon^{\mu \nu}_{(\lambda)}$ with helicity $\lambda$ can be constructed in terms of the polarization vectors of a massive vector state moving along the $z$-axis\cite{PT},
\begin{eqnarray}
\epsilon(0)^{*\mu} = (P_3, 0, 0, E)/m_T, \hspace{0.5cm} \epsilon(\pm 1)^{*\mu} = (0, \mp1, +i, 0)/\sqrt{2},
\end{eqnarray}
and are given by
\begin{eqnarray}
& & \epsilon^{\mu \nu}_{(\pm 2)} \equiv \epsilon(\pm 1)^\mu \epsilon(\pm 1)^\nu, \\
& & \epsilon^{\mu \nu}_{(\pm 1)} \equiv  \sqrt{\frac{1}{2}} [ \epsilon(\pm1)^\mu \epsilon(0)^\nu + \epsilon(0)^\mu \epsilon(\pm 1)^\nu],\\
& & \epsilon^{\mu \nu}_{(0)} \equiv \sqrt{\frac{1}{6}} [ \epsilon(+1)^\mu \epsilon (-1)^\nu + \epsilon(-1)^\mu \epsilon(+1)^\nu] + \sqrt{\frac{2}{3}} \epsilon(0)^\mu \epsilon(0)^\nu.
\end{eqnarray}
The polarization tensors are symmetric and traceless, satisfying the divergence free condition $\epsilon^{\mu \nu}_{(\lambda)} P_\nu =0$ and the orthogonal condition $\epsilon^{\mu \nu}_{(\lambda)} \epsilon^*_{(\lambda^\prime) \mu \nu} = \delta_{\lambda \lambda^\prime}$.

\subsection{Decay constants}

Contrary to the vector meson case, a $^3P_2$ tensor meson with $J^{PC}=2$ cannot be produced through the local $V-A$ and tensor currents. The decay constants should be defined via the local currents involving covariant derivatives,
\begin{eqnarray}
& & \langle T(P, \lambda)| J_{\mu \nu}(0)|0 \rangle = f_T m^2_T \epsilon^*_{(\lambda)\mu \nu}, \\
& & \langle T(P, \lambda)| J^\perp_{\mu \nu \alpha} (0)|0 \rangle = -i f^\perp_T m_T ( \epsilon^*_{(\lambda)\mu \alpha} P_\nu - \epsilon^*_{\nu \alpha} P_\mu ),
\end{eqnarray}
where
\begin{eqnarray}
& & J_{\mu \nu}(0)= \frac{1}{2} ( \bar{q}_1(0) \gamma_\mu i \overleftrightarrow{D}_\nu q_2(0) + \bar{q}_1(0) \gamma_\nu i \overleftrightarrow{D}_\mu, q_2(0), \\
& & J^\perp_{\mu \nu \alpha} (0) = \bar{q}_1 (0) \sigma_{\mu \nu} i \overleftrightarrow{D}_\alpha q_2(0),
\end{eqnarray}
and $\overleftrightarrow{D}_\mu = \overrightarrow{D}_\mu - \overleftarrow{D}_\mu$ with $\overrightarrow{D}_\mu= \overrightarrow{\partial}_\mu + i g_s A^a_\mu/2$ and $\overleftarrow{D}_\mu= \overleftarrow{\partial}_\mu - i g_s A^a_\mu/2$.

The decay constants $f_T$, $f^\perp_T$ were estimated via the QCD sum rules\cite{DecayC},
\begin{eqnarray}\label{DCT}
& & f_{f_2(1270)}= 0.102 \pm 0.006, \hspace{0.5cm} f^\perp_{f_2(1270)} = 0.117 \pm 0.025, \nonumber \\
& & f_{f^\prime_2(1525)}= 0.126 \pm 0.004, \hspace{0.5cm} f^\perp_{f^\prime_2(1525)} = 0.065 \pm 0.012, \nonumber \\
& & f_{a_2(1320)}= 0.107 \pm 0.006, \hspace{0.5cm} f^\perp_{a_2(1320)} = 0.105 \pm 0.021, \nonumber \\
& & f_{K^*_2(1430)}= 0.118 \pm 0.005, \hspace{0.5cm} f^\perp_{K^*_2(1430)} = 0.077 \pm 0.014.
\end{eqnarray}
We shall take these values as inputs in our calculations.

\subsection{Form factors}
The form factors for $B \rightarrow T$ transitons are defined by\cite{Bto2ppSL,FFY}
\begin{eqnarray}
& & \langle T(P, \lambda)|\bar{q} \gamma^\mu b |B(p_B)\rangle = -i \frac{2}{m_B + m_T} \varepsilon^{\mu \nu \alpha \beta}e^*_{(\lambda)\nu} p_{B \alpha} P_\beta V(q^2), \\
& & \langle T(P, \lambda)|\bar{q} \gamma^\mu \gamma^5 b |B(p_B)\rangle = (m_B + m_T) \left [ e^{*\mu}_{(\lambda)} - \frac{e^*_{(\lambda)} \cdot p_B}{q^2} q^\mu \right ] A_1(q^2) \nonumber \\
& & \hspace{2cm} - \frac{e^*_{(\lambda)} \cdot p_B}{m_B + m_T} \left [ (p_B + P)^\mu - \frac{m^2_B - m^2_T}{q^2} q^\mu \right ] A_2(q^2) \nonumber \\
& & \hspace{2cm} + 2 m_T \frac{e^*_{(\lambda)} \cdot p_B}{q^2} q^\mu A_0 (q^2), \\
& & \langle T(P, \lambda)|\bar{q} \sigma^{\mu \nu} q_\nu b |B(p_B)\rangle = 2 T_1(q^2) \varepsilon^{\mu \nu \alpha \beta} p_{B\nu} P_\alpha e^*_{(\lambda)\beta}, \\
& & \langle T(P, \lambda)|\bar{q} \sigma^{\mu \nu} \gamma_5 q_\nu b |B(p_B)\rangle= -i T_2(q^2) \left [ (m^2_B - m^2_T) e^{*\mu}_{(\lambda)} - (e^*_{(\lambda)} \cdot p_B) (p_B + P)^\mu \right ] \nonumber \\
& & \hspace{2cm} - i T_3(q^2) ( e^*_{(\lambda)} \cdot p_B ) \left [ q^\mu - \frac{q^2}{m^2_B - m^2_T} ( p_B + P)^\mu \right ],
\end{eqnarray}
where $q_\mu = (p_B - P)_\mu$ is the momentum transferred and $e^{*\mu}_{(\lambda)} = \epsilon^{* \mu \nu}_{(\lambda)} q_\nu/m_B$. $\varepsilon^{\mu \nu \alpha \beta}$ is the total antisymmetric Levi-Civita symbol, with $\varepsilon^{0 1 2 3} =-1$. $A_1, A_2, A_0, V$ and $T_1, T_2, T_3$ are the semileptonic and penguin type form factors, respectively. The latter are only relevant for exclusive decays induced by the flavor changing neutral currents (FCNC).

Based on the LCDAs given in Ref.\cite{DAs}, the form factors for $B \rightarrow T$ decays were calculated via the LCSR in HQEFT\cite{Bto2ppSL}. We shall use these results as inputs in our calculations, as shown in TABLE \ref{tab:FFinputs}.
\begin{table}
\centering
\begin{tabular}{|c|c|c|c|c|c|}
\hline
Decays & $F$ & $q^2=m^2_{a_2}$ & $q^2=m^2_{f_2}$ & $q^2=m^2_{K^*_2}$ &  $q^2=m^2_{f^\prime_2}$ \\
\hline
$B \rightarrow a_2$ & $A_1$ & $0.191^{+0.014}_{-0.012}$ & $0.191^{+0.014}_{-0.012}$ & $0.191^{+0.014}_{-0.012}$ &  $0.191^{+0.014}_{-0.012}$ \\
\cline{2-6}
& $A_2$ & $0.152^{+0.009}_{-0.011}$ & $0.151^{+0.009}_{-0.011}$ & $0.154^{+0.009}_{-0.012}$ &  $0.156^{+0.009}_{-0.012}$ \\
\cline{2-6}
& $V$ & $0.287^{+0.019}_{-0.015}$ & $0.286^{+0.019}_{-0.015}$ & $0.290^{+0.019}_{-0.015}$ &  $0.292^{+0.019}_{-0.015}$ \\
\hline
$B \rightarrow f_2$ & $A_1$ & $0.195^{+0.013}_{-0.013}$ & $0.195^{+0.013}_{-0.013}$ & $0.195^{+0.013}_{-0.013}$ & $0.195^{+0.013}_{-0.013}$ \\
\cline{2-6}
& $A_2$ & $0.170^{+0.009}_{-0.013}$ & $0.169^{+0.009}_{-0.012}$ & $0.172^{+0.009}_{-0.013}$ & $0.174^{+0.009}_{-0.013}$ \\
\cline{2-6}
& $V$ & $0.294^{+0.018}_{-0.016}$ & $0.293^{+0.018}_{-0.016}$ & $0.297^{+0.019}_{-0.016}$ & $0.300^{+0.019}_{-0.016}$ \\
\hline
$B \rightarrow K^*_2$ & $A_1$ & $0.180^{+0.014}_{-0.012}$ & $0.180^{+0.014}_{-0.012}$ & $0.180^{+0.014}_{-0.012}$ & $0.180^{+0.014}_{-0.012}$ \\
\cline{2-6}
& $A_2$ & $0.107^{+0.009}_{-0.008}$ & $0.107^{+0.009}_{-0.008}$ & $0.109^{+0.009}_{-0.008}$ & $0.110^{+0.009}_{-0.009}$ \\
\cline{2-6}
& $V$ & $0.263^{+0.019}_{-0.012}$ & $0.262^{+0.019}_{-0.012}$ & $0.266^{+0.019}_{-0.012}$ & $0.268^{+0.019}_{-0.013}$ \\
\hline
\end{tabular}
\caption{$B \rightarrow T$ form factors at $q^2 = m^2_{a_2}, m^2_{f_2}, m^2_{K^*_2}, m^2_{f^\prime_2}$, used as inputs\cite{Bto2ppSL}.}\label{tab:FFinputs}
\end{table}

\subsection{LCDAs}

The LCDAs of the tensor meson are defined as\cite{Bto2ppSL,DAs}
\begin{eqnarray}
& & \langle T(P, \lambda)| \bar{q}_1 (x) \gamma_\mu q_2(0)|0 \rangle = f_T m^2_T \int^1_0 du e^{iuP\cdot x} \left [ P_\mu \frac{\epsilon^*_{(\lambda)\alpha \beta} x^\alpha x^\beta}{(P \cdot x)^2} \phi_\parallel (u) + \left ( \frac{\epsilon^*_{(\lambda)\mu \alpha} x^\alpha }{P \cdot x } \right.  \right. \nonumber \\
& & \hspace{2cm} \left. \left. -P_\mu \frac{\epsilon^*_{(\lambda)\alpha \beta} x^\alpha x^\beta}{(P \cdot x)^2} \right ) g_v (u) - \frac{1}{2} x_\mu \frac{\epsilon^*_{(\lambda)\alpha \beta} x^\alpha x^\beta}{(P \cdot x)^3} m^2_T ( g_3(u) + \phi_\parallel (u) -2 g_v (u)) \right ], \\
& & \langle T(P, \lambda)| \bar{q}_1 (x) \gamma_\mu \gamma_5 q_2(0)|0 \rangle = \frac{1}{2} f_T m^2_T \int^1_0 d u e^{iuP \cdot x} \varepsilon_{\mu \nu \alpha \beta} x^\nu P^\alpha \epsilon^{* \beta \delta}_{(\lambda)} x_\delta \frac{1}{P \cdot x} g_a (u), \\
& & \langle T(P, \lambda)| \bar{q}_1 (x) \sigma_{\mu \nu} q_2(0)|0 \rangle = -i f^\perp_T m_T \int^1_0 d u e^{i u P \cdot x} \left [ \left ( \epsilon^*_{(\lambda) \mu \alpha} x^\alpha P_\nu - \epsilon^*_{(\lambda) \nu \alpha} x^\alpha P_\mu \right ) \right. \nonumber \\
& & \hspace{2cm} \times \frac{1}{P \cdot x} \phi_\perp (u) + ( P_\mu x_\nu - P_\nu x_\mu ) \frac{m^2_T \epsilon^*_{(\lambda)\alpha \beta} x^\alpha x^\beta}{(P \cdot x)^3} \left ( h_t (u) - \frac{1}{2} \phi_\perp (u) - \frac{1}{2} h_3(u) \right ) \nonumber \\
& & \hspace{2cm} \left. +\frac{1}{2} \left ( \epsilon^*_{(\lambda) \mu \alpha} x^\alpha x_\nu - \epsilon^*_{(\lambda) \nu \alpha} x^\alpha x_\mu \right ) \frac{m^2_T}{(P \cdot x)^2} ( h_3(u) - \phi_\perp (u)) \right ], \\
& & \langle T(P, \lambda)| \bar{q}_1 (x) q_2(0)|0 \rangle = - \frac{i}{2} f^\perp_T m^3_T \int^1_0 d u e^{iu P \cdot x} \frac{\epsilon^*_{(\lambda) \alpha \beta} x^\alpha x^\beta}{ P \cdot x} h_s (u),
\end{eqnarray}
where $\{\phi_\parallel (u), \phi_\perp (u) \}$, $\{ g_v (u), g_a (u), h_t (u), h_s (u) \}$ and $\{g_3(u), h_3(u)\}$ are of twist-2, twist-3 and twist-4, respectively.

The asymptotic twist-2 and twist-3 LCDAs of tensor mesons are given by\cite{DAs}
\begin{eqnarray}
& & \phi_{\parallel(\perp)} (u) = 30u (1-u) (2u-1), \label{DAsT1}\\
& & g_v(u) =5 (2u-1)^2, \label{DAsT2}\\
& & g_a(u) = 10 u (1-u)(2u-1), \label{DAsT3}\\
& & h_t(u) = \frac{15}{2} (2u-1)(1-6u +6 u^2), \label{DAsT4}\\
& & h_s(u) = 15u (1-u) (2u-1).\label{DAsT5}
\end{eqnarray}
It is also useful to introduce\cite{QCDF}
\begin{eqnarray}
& & \phi_+ (u) = \int^1_u dv \frac{\phi_\parallel (v)}{v}, \label{DAsT6}\\
& & \phi_-(u)= \int^u_0 d v \frac{\phi_\parallel(v)}{1-v}, \label{DAsT6}\\
& & \phi_t(u) = \int^u_0 d v \frac{\phi_\perp (v)}{1-v} - \int^1_u d v \frac{\phi_\perp (v)}{ v}.\label{DAsT8}
\end{eqnarray}
We shall use these expressions as inputs in our calculations.

\section{$B \rightarrow TT$ decays}\label{SecIII}

Due to the fact that the longitudinal and transverse helicity projection operators for the tensor meson are very similar to the projectors for the vector meson\cite{QCDF}, the nonfactorizable contributions can be obtained directly from the $B \rightarrow TV$ decays by making some suitable replacements.

Within the framework of QCD factorization\cite{QCDFO1,QCDFO2,QCDFO3}, the effective Hamiltonian matrix elements can be written in the form
\begin{eqnarray}
\langle T_1 T_2 | {\cal H}_{eff} | B \rangle = \frac{G_F}{\sqrt{2}} \sum_{p=u,c} \lambda^{(q)}_p \langle T_1 T_2 | {\cal T}^{p,h}_{\cal A} + {\cal T}^{p,h}_{\cal B} | B \rangle,
\end{eqnarray}
where $\lambda^{(q)}_p \equiv V_{pb} V^*_{pq}$ with $q= s, d$. In the rest frame of $B$ meson, since it has spin zero, the helicities of $T_1$ and $T_2$ are equal, denoted as $h$ here. ${\cal T}^{p,h}_{\cal A}$ contains the contributions from naive factorization, vertex corrections, penguin contractions and spectator scattering, while ${\cal T}^{p,h}_{\cal B}$ describes the contributions from annihilation topology amplitudes. In general, the ${\cal T}^{p,h}_{\cal A}$ part can be expressed in terms of $c \alpha_i^{p,h} (T_1 T_2) X^{(B T_1, T_2)}_h$, where $c$ contains the factors arising from the flavor structures of final state mesons, $\alpha_i^{p, h}(T_1 T_2)$ is the coefficients of the flavor operators and $X^{(B T_1 T_2)}_h$ the factorizable amplitudes.

The helicity of tensor meson can be $h=0, \pm1$. (Note that the $h=\pm 2$ configurations are not allowed physically\cite{pQCDBs}.) In the rest frame of $B$ meson, the longitudinal ($h=0$) and transverse ($h= \pm 1)$ components of factorizable amplitudes can be written as
\begin{eqnarray}
& & X^{(B T_1, T_2)}_0 = \frac{i f_{T_2}}{2m^2_{T_1}} p_c \sqrt{\frac{2}{3}} \left [ (m^2_B -m^2_{T_2} - m^2_{T_1})(m_B + m_{T_1}) A^{BT_1}_1 (m^2_{T_2}) \right. \nonumber \\
& & \hspace{2cm} \left. - \frac{4 m^2_B p^2_c}{m_B + m_{T_1}} A^{BT_1}_2 (m^2_{T_2}) \right ], \\
& & X^{(B T_1, T_2)}_\pm = -i f_{T_2} m_B m_{T_2} \frac{p_c}{\sqrt{2}m_{T_1}} \left [ \left ( 1+ \frac{m_{T_1}}{m_B} \right ) A^{BT_1}_1 (m^2_{T_2}) \right. \nonumber \\
& & \hspace{2cm} \left. \mp \frac{2p_c}{m_B + m_{T_1}} V^{BT_1} (m^2_{T_2}) \right ],
\end{eqnarray}
where $p_c$ is the magnitude of center of mass momentum of final state meson.

The coefficients of flavor operators $\alpha_i^{p, h}(T_1 T_2)$ can be expressed in terms of the coefficients of factorized operators $a_i^{p, h}(T_1 T_2)$ as\cite{QCDFO4}
\begin{eqnarray}
& & \alpha^h_1(T_1T_2)= a^h_1(T_1 T_2), \\
& & \alpha^h_2(T_1T_2)= a^h_2(T_1 T_2), \\
& & \alpha^{p, h}_3(T_1T_2)= a^{p,h}_3(T_1 T_2) + a^{p,h}_5(T_1 T_2), \\
& & \alpha^{p, h}_4(T_1T_2)= a^{p,h}_4(T_1 T_2) - r^{T_2}_\chi a^{p,h}_6(T_1 T_2), \\
& & \alpha^{p, h}_{3, EW}(T_1T_2)= a^{p,h}_9(T_1 T_2) + a^{p,h}_7(T_1 T_2), \\
& & \alpha^{p, h}_{4, EW}(T_1T_2)= a^{p,h}_{10}(T_1 T_2) - r^{T_2}_\chi a^{p,h}_8(T_1 T_2),
\end{eqnarray}
where
\begin{eqnarray}
r^T_\chi (\mu) = \frac{2 m_T}{m_b(\mu)} \frac{f^\perp_T (\mu)}{ f_T}
\end{eqnarray}
is the chiral factor.

The coefficients of factorized operators $a_i^{p, h}(T_1 T_2)$ are basically the Wilson coefficients in conjunction with short distance nonfactorizable corrections such as vertex corrections and hard spectator interactions. In general, they have the expressions\cite{QCDF,QCDFO4}
\begin{eqnarray}
a^{p,h}_i (T_1 T_2) = \left ( c_i + \frac{c_{i \pm 1}}{N_c} \right ) N^h_i (T_2) + \frac{c_{i \pm 1}}{N_c} \frac{C_F \alpha_s}{4 \pi} \left [ V^h_i (T_2) + \frac{4 \pi^2}{N_c} H^h_i ( T_1 T_2) \right ] + P^{p,h}_i (T_2),
\end{eqnarray}
where $i =1, \cdots, 10$, the upper (lower) signs apply when $i$ is odd (even), $c_i$ are the Wilson coefficients. $C_F = (N^2_c -1)/(2 N_c)$, with $N_c=3$. $T_2$ is the emitted meson and $T_1$ shares the same spectator quark with the $B$ meson. The quantities $N^h_i (T_2)$, $V^h_i (T_2)$, $H^h_i ( T_1 T_2)$ and $P^{p,h}_i (T_2)$ account for the contributions from naive factorization, vertex corrections, hard spectator interactions and penguin contractions, respectively. For tensor meson, $N^h_i(T_2)=0$. The expression for $V^h_i (T_2)$, $H^h_i ( T_1 T_2)$ and $P^{p,h}_i (T_2)$ are as following.

The vertex corrections are given by
\begin{eqnarray}
& & V^0_i (T_2) = \left\{
           \begin{array}{ll}
                C(T_2) \int^1_0 d x \phi_\parallel (x) \left [ 12 \log \frac{m_b}{\mu} -18 +g(x) \right ], & (i=1-4, 9, 10),  \\
              C(T_2) \int^1_0 d x \phi_\parallel (x) \left [ -12 \log \frac{m_b}{\mu} +6 -g(1-x) \right ], & (i=5, 7), \\
              \frac{1}{C(T_2)} \int^1_0 d x \phi_t(x) \left [ -6 + h(x) \right ], & (i=6, 8).
           \end{array}
         \right. \\
& & V^\pm_i (T_2) = \left\{
           \begin{array}{ll}
                D(T_2) \int^1_0 d x \phi_\pm (x) \left [ 12 \log \frac{m_b}{\mu} -18 +g_T(x) \right ], & (i=1-4, 9, 10),  \\
              D(T_2) \int^1_0 d x \phi_\mp (x) \left [ -12 \log \frac{m_b}{\mu} +6 -g_T(1-x) \right ], & (i=5, 7), \\
              0, & (i=6, 8).
           \end{array}
         \right.
\end{eqnarray}
with
\begin{eqnarray}
& & g(x) = 3 \left ( \frac{1-2x}{1-x} \ln x - i \pi \right ) + \left [ 2 {\rm Li}_2 (x) + \frac{2 \ln x}{1-x} - (3+2i \pi) \ln x - (x \leftrightarrow 1-x ) \right ], \\
& & h(x) = 2 {\rm Li}_2 (x) - \ln^2 x - (1+ 2 \pi i ) \ln x - (x \leftrightarrow 1-x ), \\
& & g_T (x) = g(x) + \frac{\ln x}{1-x}.
\end{eqnarray}
For tensor meson, $C(T) = \sqrt{\frac{2}{3}}$, $D(T) = \frac{1}{\sqrt{2}}$.

$H^h_i ( T_1 T_2)$ arise from hard spectator interactions with a hard gluon exchange between the emitted meson and the spectator quark of the $B$ meson. $H^0_i (T_1 T_2)$ have the expressions
\begin{eqnarray}
& & H^0_i( T_1 T_2) = \frac{i f_B f_{T_1} f_{T_2}}{X^{(B T_1, T_2)}_0} \frac{m_B}{\lambda_B} C(T_2) \nonumber \\
& & \hspace{2cm} \times \left\{
           \begin{array}{ll}
                \int^1_0 d u d v  \left [ C(T_1) \frac{\phi_\parallel (u) \phi_\parallel (v)}{ (1-u) (1-v)} + r^{T_1}_\chi \frac{\phi_t (u) \phi_\parallel (v)}{ C(T_1) (1-u) v } \right ], & (i=1-4 , 9, 10),  \\
               \int^1_0 d u d v  \left [ C(T_1) \frac{\phi_\parallel (u) \phi_\parallel (v)}{ (1-u) v} + r^{T_1}_\chi \frac{\phi_t (u) \phi_\parallel (v)}{ C(T_1) (1-u) (1- v) } \right ], & (i=5, 7), \\
                             0, & (i=6, 8).
           \end{array}
         \right.
\end{eqnarray}
The transverse hard spectator terms $H^\pm_i (T_1 T_2)$ read
\begin{eqnarray}
& & H^+_i (T_1 T_2) = - \frac{\sqrt{2} i f_B f_{T_1} f_{T_2} m_{T_1} m_{T_2}}{m^2_B X^{(B T_1, T_2)}_+} \frac{m_B}{\lambda_B} \left\{
           \begin{array}{ll}
                \int^1_0 d u d v  \frac{(1-u -v) \phi_+ (u) \phi_+ (v)}{ (1-u)^2 ( 1-v)^2 }, & (i=1-4 , 9, 10),  \\
               \int^1_0 d u d v  \frac{( u-v) \phi_+ (u) \phi_- (v)}{ (1-u)^2 v^2}, & (i=5, 7), \\
                             0, & (i=6, 8).
            \end{array}
         \right. \\
& & H^-_i (T_1 T_2) =  \frac{\sqrt{2} i f_B f^\perp_{T_1} f_{T_2}  m_{T_2}}{m_B X^{(B T_1, T_2)}_-} \frac{m_B}{\lambda_B} \left\{
           \begin{array}{ll}
                \int^1_0 d u d v  \frac{ \phi_\perp (u) \phi_\perp (v)}{ (1-u)^2 v }, & (i=1-4 , 9, 10),  \\
               \int^1_0 d u d v  \frac{ - \phi_\perp (u) \phi_+ (v)}{ (1-u)^2 (1-v)}, & (i=5, 7), \\
               \frac{f_{T_1}m_B m_{T_1}}{2 f^\perp_{T_1} m^2_{T_2}} \int^1_0 du dv \frac{- \phi_+ (u) \phi_\perp (v)}{v (1-u) (1-v)}, & (i=6, 8).
            \end{array}
         \right.
\end{eqnarray}

At order $\alpha_s$, corrections from penguin contractions are present only for $i=4, 6$. For $i=4$, we have
\begin{eqnarray}
& & P^{p, h}_4 (T_2) = \frac{C_F \alpha_s}{4 \pi N_c} \left \{ c_1 \left [ G^h_{T_2} (s_p) + g^h_{T_2} \right ] + c_3 \left [ G^h_{T_2} ( s_s) + G^h_{T_2} (1) + 2 g^h_{T_2} \right ]  \right. \nonumber \\
& & \hspace{2cm} \left. +(c_4 +c_6) \sum_{i=u}^b \left [ G^h_{T_2} (s_i) + g^{\prime h}_{T_2} \right ] - 2 c^{eff}_{8g} G^h_g  \right \},
\end{eqnarray}
where $s_i = m^2_i/m^2_b$ and
\begin{eqnarray}
& & G^h_{T_2} (s)= 4 \int^1_0 d u \phi^{T_2, h} (u) \int^1_0 d x x (1-x) \ln [s - (1-u) x (1-x) - i \epsilon ],  \\
& & g^h_{T_2} = \left ( \frac{4}{3} \ln \frac{m_b}{\mu} + \frac{2}{3} \right ) \int^1_0 d x \phi^{T_2, h} (x), \\
& & g^{\prime h}_{T_2} = \frac{4}{3} \ln \frac{m_b}{\mu} \int^1_0 d x \phi^{T_2, h} (x),
\end{eqnarray}
with $\phi^{T_2, 0} = C(T_2) \phi_\parallel$, $\phi^{T_2, \pm} = D(T_2) \phi_\pm$.
For $i=6$, the results for penguin contribution is
\begin{eqnarray}
P^{p, h}_6 (T_2) = \frac{G_F \alpha_s}{ 4 \pi N_c} \left \{ c_1 \hat{G}^h_{T_2} (s_p) + c_3 \left [ \hat{G}^h_{T_2} (s_s) + \hat{G}^h_{T_2} (1) \right ] + (c_4 +c_6) \sum_{i=u}^b \hat{G}^h_{T_2}( s_i) \right \},
\end{eqnarray}
where
\begin{eqnarray}
& & \hat{G}^0_{T_2} (s) = \frac{4}{C(T_2)} \int^1_0 d u \phi_t (u) \int^1_0 d x x (1-x) \ln [s - (1-u) x (1-x) - i \epsilon ] , \\
& & \hat{G}^\pm_{T_2} (s) =0.
\end{eqnarray}
Including the electromagnetic corrections, for $i=8, 10$, we find
\begin{eqnarray}
& & P^{p, h}_8 (T_2) = \frac{\alpha}{ 9 \pi N_c} (c_1 + N_c c_2 ) \hat{G}^h_{T_2} (s_p ), \\
& & P^{p, h}_{10} (T_2) =  \frac{\alpha}{ 9 \pi N_c} \left \{ (c_1 + N_c c_2) \left [ G^h_{T_2} (s_p) + 2 g_{T_2} \right ] - 3 c^{eff}_{7 \gamma} G^h_g \right \},
\end{eqnarray}
where
\begin{eqnarray}
& & G^0_g = C(T_2) \int^1_0 d u \frac{\phi_\parallel (u)}{1-u}, \\
& & G^\pm_g =0.
\end{eqnarray}

The weak annihilation contributions to the $B \rightarrow T_1 T_2$ decays can be described in terms of the building blocks $b^{p, h}_i$ and $b^{p, h}_{i EW}$,
\begin{eqnarray}
\langle T_1 T_2 | {\cal T}^{p, h}_B | B \rangle = i f_B f_{T_1} f_{T_2} \sum_i \left ( d_i b^{p, h}_i + d^\prime_i b^{p, h}_{i, EW} \right ).
\end{eqnarray}
The building blocks have the expressions
\begin{eqnarray}
& & b^{p, h}_1= \frac{C_F}{N^2_c} c_1 A^{i, h}_1, \\
& & b^{p, h}_2 = \frac{C_F}{N^2_c} c_2 A^{i, h}_1, \\
& & b^{p, h}_3= \frac{C_F}{N^2_c} \left [ c_3 A^{i, h}_1 +c_5 (A^{i, h}_3 + A^{f, h}_3) + N_c c_6 A^{f,h}_3 \right ], \\
& & b^{p, h}_4= \frac{C_F}{N^2_c} \left [ c_4 A^{i, h}_1 + c_6 A^{f, h}_2 \right ], \\
& & b^{p, h}_{3, EW}= \frac{C_F}{N^2_c} \left [ c_9 A^{i, h}_1 +c_7 (A^{i, h}_3 + A^{f, h}_3) + N_c c_8 A^{f, h}_3 \right ], \\
& & b^{p, h}_{4, EW}= \frac{C_F}{N^2_c} \left [ c_{10} A^{i, h}_1 + c_8 A^{f, h}_2 \right ].
\end{eqnarray}
The subscripts $1$, $2$ and $3$ of $A^{i(f), h}_n$ denote the annihilation amplitudes induced by $(V-A)(V-A)$, $(V-A)(V+A)$ and $(S-P)(S+P)$ operators, respectively, and the superscripts $i(f)$ refer to gluon emission from the initial (final) state quarks.
Choosing the convention that $T_1$ contains a quark from the weak vertex with momentum fraction $u$ and $T_2$ an antiquark with momentum fraction $1-v$, the weak annihilation amplitudes
\begin{eqnarray}
& & A^{i, 0}_1 (T_1T_2) = \sqrt{\frac{2}{3}} \pi \alpha_s \int^1_0 d u d v \left \{ \phi_\parallel (u) \phi_\parallel (v) \left [ \frac{1}{u (1- (1-u)v)} + \frac{1}{u (1-v)^2 } \right ] \right. \nonumber \\
& & \hspace{2.2cm} \left. \mp \frac{3}{2} r^{T_1}_\chi r^{T_2}_\chi \phi_t (u) \phi_t (v) \frac{2}{u (1-v)} \right \},  \\
& & A^{i, +}_1 (T_1 T_2) = - \pi \alpha_s \frac{\sqrt{2} m_{T_1} m_{T_2}}{ m^2_B} \int^1_0 d u d v \left \{ \phi_+ (u) \phi_+ (v) \left [ \frac{2}{u (1-v)^3 } \right. \right. \nonumber \\
& & \hspace{2.2cm} \left. \left. - \frac{v}{(1- (1-u)v)^2} - \frac{v}{ (1-v)^2 (1- (1-u)v)} \right ] \right \}, \\
& & A^{i, -}_1 (T_1 T_2) = - \pi \alpha_s \frac{\sqrt{2} m_{T_1} m_{T_2}}{ m^2_B} \int^1_0 d u d v \left \{ \phi_- (u) \phi_- (v) \left [ \frac{(1-u)+ (1-v)}{u^2 (1-v)^2 } \right. \right. \nonumber \\
& & \hspace{2.2cm} \left. \left. + \frac{1}{(1- (1-u) v )^2 } \right ] \right \}, \\
& & A^{i, 0}_2 (T_1T_2) = \sqrt{\frac{2}{3}} \pi \alpha_s \int^1_0 d u d v \left \{ \phi_\parallel (u) \phi_\parallel (v) \left [ \frac{1}{(1-v) (1- (1-u)v)} + \frac{1}{u^2 (1-v) } \right ] \right. \nonumber \\
& & \hspace{2.2cm} \left. \mp \frac{3}{2} r^{T_1}_\chi r^{T_2}_\chi \phi_t (u) \phi_t (v) \frac{2}{u (1-v)} \right \},\\
& & A^{i, +}_2 (T_1 T_2) = - \pi \alpha_s \frac{\sqrt{2} m_{T_1} m_{T_2}}{ m^2_B} \int^1_0 d u d v \left \{ \phi_- (u) \phi_- (v) \left [ \frac{2}{u^3 (1-v) } \right. \right. \nonumber \\
& & \hspace{2.2cm} \left. \left. - \frac{1-u}{(1- (1-u)v)^2} - \frac{1-u}{ u^2 (1- (1-u)v)} \right ] \right \}, \\
& & A^{i, -}_2 (T_1 T_2) = - \pi \alpha_s \frac{\sqrt{2} m_{T_1} m_{T_2}}{ m^2_B} \int^1_0 d u d v \left \{ \phi_+ (u) \phi_+ (v) \left [ \frac{u+ v}{u^2 (1-v)^2 } \right. \right. \nonumber \\
& & \hspace{2.2cm} \left. \left. + \frac{1}{(1- (1-u) v )^2 } \right ] \right \}, \\
& & A^{i, 0}_3 (T_1 T_2) = \pi \alpha_s \int^1_0 d u d v \left \{ \frac{C(T_2)}{C(T_1)} r^{T_1}_\chi \phi_t (u) \phi_\parallel (v) \frac{2 (1-u)}{ u (1-v) (1- (1-u)v )} \right. \nonumber \\
& & \hspace{2.2cm} \left. + \frac{C(T_1)}{C(T_2)} r^{T_2}_\chi \phi_\parallel (u) \phi_t (v) \frac{2v}{u (1-v) (1-(1-u)v)} \right \}, \\
& & A^{i, -}_3 (T_1 T_2) = - \frac{\pi \alpha_s}{\sqrt{2}} \int^1_0 d u d v \left \{ - \frac{m_{T_2}}{m_{T_1}} r^{T_1}_\chi \phi_\perp (u) \phi_- (v) \frac{2}{ u (1-v) (1- (1-u)v)} \right. \nonumber \\
& & \hspace{2.2cm} \left. + \frac{m_{T_1}}{m_{T_2}} r^{T_2}_\chi \phi_+ (u) \phi_\perp (v) \frac{2}{ u (1-v) (1-(1-u)v)} \right \}, \\
& &  A^{f, 0}_3 (T_1 T_2) = \pi \alpha_s \int^1_0 d u d v \left \{ \frac{C(T_2)}{C(T_1)} r^{T_1}_\chi \phi_t (u) \phi_\parallel (v) \frac{2 (1+(1-v))}{ u (1-v)^2} \right. \nonumber \\
& & \hspace{2.2cm} \left. - \frac{C(T_1)}{C(T_2)} r^{T_2}_\chi \phi_\parallel (u) \phi_t (v) \frac{2(1+u)}{u^2 (1-v)} \right \}, \\
& & A^{f, -}_3 (T_1 T_2) = - \frac{\pi \alpha_s}{\sqrt{2}} \int^1_0 d u d v \left \{  \frac{m_{T_2}}{m_{T_1}} r^{T_1}_\chi \phi_\perp (u) \phi_- (v) \frac{2}{ u^2 (1-v) } \right. \nonumber \\
& & \hspace{2.2cm} \left. + \frac{m_{T_1}}{m_{T_2}} r^{T_2}_\chi \phi_+ (u) \phi_\perp (v) \frac{2}{ u (1-v)^2} \right \},
\end{eqnarray}
and $A^{f, h}_1 = A^{f, h}_2 = A^{i, +}_3 = A^{f, +}_3 =0$. Since the annihilation contributions $A^{i, \pm}_{1, 2}$ are suppressed by a factor of $m_{T_1} m_{T_2}/m^2_B$ relative to other terms, in the numerical analysis we will consider only the annihilation contributions due to $A^{i, 0}_{1, 2, 3}$, $A^{i, -}_3$, $A^{f, 0}_3$ and $A^{f, -}_3$.

Although the coefficients of the flavor operators $\alpha^{p, h}_i$ are formally renormalization scale and $\gamma_5$ scheme independent, in practice there exists some residual scale dependence to finite order. To be specific, we shall evaluate the vertex corrections to the decay amplitude at the scale $\mu =m_b$, while the hard spectator and annihilation contributions at the hard collinear scale $\mu_h = \sqrt{\mu \Lambda_h}$ with $\Lambda_h \approx 0.5{\rm GeV}$\cite{QCDF}. In addition, power corrections in QCD factorization always involve troublesome endpoint divergences. Since the treatment of endpoint divergences is model dependent, subleading power corrections generally can be studied only in a phenomenological way. We shall model the endpoint divergence $X \equiv \int^1_0 d x /(1-x)$ in the annihilation and hard spectator scattering diagrames as\cite{QCDF,QCDFO1,QCDFO2,QCDFO3,QCDFO4}
\begin{eqnarray}
& & X_A  = \ln \left ( \frac{m_B}{\Lambda_h} \right ) ( 1 + \rho_A e^{i \phi_A} ),\label{XA} \\
& & X_H  = \ln \left ( \frac{m_B}{\Lambda_h} \right ) ( 1 + \rho_H e^{i \phi_H} ),\label{XH}
\end{eqnarray}
with the unknown real parameters $\rho_{A, H}$ and $\phi_{A, H}$. For simplicity, we shall assume that $X_{A, H}$ are helicity independent.

Considering the asymptotic distribution amplitudes for $\phi_{\parallel (\perp)}$, $\phi_{+(-)}$ and $\phi_t$, we find
\begin{eqnarray}
& & A^{i, 0}_1 = 50 \sqrt{\frac{2}{3}} \pi \alpha_s \left [ 3 (71-7\pi^2 -X_A) - \frac{3}{2} r^{T_1}_\chi r^{T_2}_\chi ( -3 + X_A)^2 \right ], \\
& & A^{i, 0}_3 = 50 \pi \alpha_s \left ( 120 -11 \pi^2 -12 X_A +3 X^2_A \right ) \left ( r^{T_1}_\chi - r^{T_2}_\chi \right ), \\
& & A^{i, -}_3 = 300 \frac{\pi \alpha_s}{\sqrt{2}} \left ( 12 - \pi^2 - 2X_A + \frac{1}{2} X^2_A \right ) \left ( - \frac{m_{T_2}}{m_{T_1}} r^{T_1}_\chi + \frac{m_{T_1}}{m_{T_2}} r^{T_2}_\chi \right ), \\
& & A^{f, 0}_3 = 50 \pi \alpha_s \left ( -3 + X_A \right ) \left ( -11 + 6 X_A \right ) \left ( r^{T_1}_\chi + r^{T_2}_\chi \right ), \\
& & A^{f, -}_3 = 50 \frac{\pi \alpha_s}{\sqrt{s}} \left (2 -X_A \right ) \left ( 17- 6 X_A \right ) \left ( \frac{m_{T_2}}{m_{T_1}} r^{T_1}_\chi + \frac{m_{T_1}}{m_{T_2}} r^{T_2}_\chi \right ),
\end{eqnarray}
and $A^{i, 0}_2 = A^{i, 0}_1$.

By using the trace formalism over flavor matrices\cite{QCDFO4}, we obtain the explicit expressions for the decay amplitudes of all relevant $B \rightarrow TT$ modes as follows,
\begin{eqnarray}
& & {\cal A}_{B^- \rightarrow a^-_2 a^0_2} = \frac{G_F}{2} \sum_{p=u,c} \lambda^{(d)}_p  \left [ \delta_{p u} \left ( \alpha^h_1 + \alpha^h_2 \right ) + \frac{3}{2} \alpha^{p, h}_{3, EW} + \frac{3}{2} \alpha^{p, h}_{4, EW} \right ] X^{(B a_2, a_2 )}_h, \\
& & {\cal A}_{B^0 \rightarrow a^-_2 a^+_2} =  \frac{G_F}{\sqrt{2}} \sum_{p=u,c} \lambda^{(d)}_p \left [ \delta_{p u} \left ( \alpha^h_1 + \beta^h_1 \right ) + \alpha^{p,h}_4 + \alpha^{p, h}_{4, EW} + \beta^{p, h}_3 +2 \beta^{p, h}_4 \right. \nonumber \\
& & \hspace{2.2cm} \left. - \frac{1}{2} \beta^{p, h}_{3, EW} + \frac{1}{2} \beta^{p, h}_{4, EW} \right ] X^{(B a_2, a_2)}_h, \\
& & {\cal A}_{B^0 \rightarrow a^0_2 a^0_2} = - \frac{G_F}{2\sqrt{2}} \sum_{p=u,c} \lambda^{(d)}_p \left [ \delta_{pu} \left ( \alpha^h_2 - \beta^h_1 \right ) - \alpha^{p, h}_4 + \frac{3}{2} \alpha^{p, h}_{3, EW} + \frac{1}{2} \alpha^{p, h}_{4, EW} \right. \nonumber \\
& & \hspace{2.2cm} \left. - \beta^{p, h}_3 - 2 \beta^{p, h}_4 + \frac{1}{2} \beta^{p, h}_{3, EW} - \frac{1}{2} \beta^{p, h}_{4, EW} \right ] X^{(B a_2, a_2)}_h, \\
& & {\cal A}_{B^- \rightarrow a^-_2 f_2} = \frac{G_F}{\sqrt{2}} \sum_{p=u,c} \lambda^{(d)}_p \left \{ \frac{1}{\sqrt{2}} \left [ \delta_{pu} \left ( \alpha^h_2 + \beta^h_2 \right ) +2 \alpha^{p, h}_3 + \alpha^{p, h}_4 + \frac{1}{2} \alpha^{p, h}_{3, EW} \right. \right. \nonumber \\
& & \hspace{2.2cm} \left. -\frac{1}{2} \alpha^{p, h}_{4, EW} + \beta^{p, h}_3 + \beta^{p, h}_{3, EW} \right ] X^{(B a_2, f_2 )}_h + \left [ \delta_{pu} \left ( \alpha^h_1 + \beta^h_2 \right ) \right. \nonumber \\
& & \hspace{2.2cm} \left. \left. + \alpha^{p, h}_4 + \alpha^{p, h}_{4, EW} + \beta^{p, h}_3 + \beta^{p, h}_{3, EW} \right ] X^{(B f_2, a_2)}_h \right \}, \\
& & {\cal A}_{B^0 \rightarrow a^0_2 f_2} =  \frac{G_F}{ \sqrt{2}} \sum_{p=u,c} \lambda^{(d)}_p \left \{ -\frac{1}{2} \left [ \delta_{p u} \left ( \alpha^h_2 - \beta^h_1 \right ) + 2 \alpha^{p, h}_3 + \alpha^{p, h}_4 + \frac{1}{2} \alpha^{p, h}_{3, EW} \right. \right. \nonumber \\
& & \hspace{2.2cm} \left. - \frac{1}{2} \alpha^{p, h}_{4, EW} + \beta^{p, h}_3 - \frac{1}{2} \beta^{p, h}_{3, EW} - \frac{3}{2} \beta^{p, h}_{4, EW} \right ] X^{(B a_2, f_2)}_h \nonumber \\
& & \hspace{2.2cm} - \frac{1}{\sqrt{2}} \left [ - \delta_{p u} \left ( \alpha^h_2 + \beta^h_1 \right ) + \alpha^{p, h}_4 - \frac{3}{2} \alpha^{p, h}_{3,EW} - \frac{1}{2} \alpha^{p, h}_{4, EW} \right. \nonumber \\
& & \hspace{2.2cm} \left. \left. + \beta^{p, h}_3 - \frac{1}{2} \beta^{p, h}_{3, EW} - \frac{3}{2} \beta^{p, h}_{4, EW} \right ] X^{(Bf_2, a_2)}_h \right \}, \\
& & {\cal A}_{B^- \rightarrow a^-_2 f^\prime_2} = \frac{G_F}{\sqrt{2}} \sum_{p=u,c} \lambda^{(d)}_p \left ( \alpha^{p, h}_3 - \frac{1}{2} \alpha^{p, h}_{3, EW} \right ) X^{(B a_2, f^\prime_2)}_h, \\
& & {\cal A}_{B^0 \rightarrow a^0_2 f^\prime_2} = - \frac{G_F}{2} \sum_{p=u,c} \lambda^{(d)}_p \left ( \alpha^{p, h}_3 - \frac{1}{2} \alpha^{p, h}_{3, EW} \right ) X^{(B a_2, f^\prime_2)}_h, \\
& & {\cal A}_{B^0 \rightarrow f_2 f_2} =  \frac{G_F}{2} \sum_{p=u,c} \lambda^{(d)}_p \left [ \delta_{pu} \left ( \alpha^h_2 + \beta^h_1 \right ) +2 \alpha^{p, h}_3 + \alpha^{p, h}_4 + \frac{1}{2} \alpha^{p, h}_{3, EW} - \frac{1}{2} \alpha^{p, h}_{4, EW} \right. \nonumber \\
& & \hspace{2.2cm} \left. + \beta^{p, h}_3 + 2 \beta^{p, h}_4 - \frac{1}{2} \beta^{p, h}_{3, EW} + \frac{1}{2} \beta^{p, h}_{4, EW} \right ] X^{(B f_2, f_2)}_h, \\
& & {\cal A}_{B^0 \rightarrow f_2 f^\prime_2} =  \frac{G_F}{2 } \sum_{p=u,c} \lambda^{(d)}_p \left ( \alpha^{p, h}_3 - \frac{1}{2} \alpha^{p, h}_{3, EW} \right ) X^{(B f_2, f^\prime_2)}_h, \\
& & {\cal A}_{B^- \rightarrow K^{*-}_2 K^{*0}_2} = \frac{G_F}{\sqrt{2}} \sum_{p=u,c} \lambda^{(d)}_p \left ( \delta_{pu} \beta^h_2 + \alpha^{p, h}_4 - \frac{1}{2} \alpha^{p, h}_{4, EW} + \beta^{p, h}_3 + \beta^{p, h}_{3, EW} \right ) X^{(B K^*_2, K^*_2)}_h, \\
& & {\cal A}_{B^0 \rightarrow K^{*-}_2 K^{*+}_2} = \frac{G_F}{\sqrt{2}} \sum_{p=u,c} \lambda^{(d)}_p \left [ \left (  \delta_{pu} \beta^h_1 + \beta^{p, h}_4  + \beta^{p,h}_{4, EW} \right )  X^{(B K^*_2, K^*_2)}_h  \right. \nonumber \\
& & \hspace{2.7cm} \left.  + i f_B f^2_{K^*_2} \left ( b^{p, h}_4 - \frac{1}{2} b^{p, h}_{4, EW} \right ) \right ], \\
& & {\cal A}_{B^0 \rightarrow K^{*0}_2 K^{*0}_2} = \frac{G_F}{\sqrt{2}} \sum_{p=u,c} \lambda^{(d)}_p \left [ \left ( \alpha^{p,h}_4 - \frac{1}{2} \alpha^{p, h}_{4, EW} + \beta^{p, h}_3  - \frac{1}{2} \beta^{p, h}_{3, EW} + \beta^{p, h}_4 \right. \right. \nonumber \\
& & \hspace{2.7cm} \left. \left. - \frac{1}{2} \beta^{p, h}_{4, EW} \right ) X^{(B K^*_2, K^*_2)}_h + i f_B f^2_{K^*_2} \left ( b^{p, h}_4 - \frac{1}{2} b^{p, h}_{4, EW} \right ) \right ] , \\
& & {\cal A}_{B^- \rightarrow a^0_2 K^{*-}_2} =\frac{G_F}{2} \sum_{p=u,c} \lambda^{(s)}_p \left \{ \left [ \delta_{pu} \left ( \alpha^h_1 + \beta^h_2 \right ) + \alpha^{p, h}_4 + \alpha^{p, h}_{4, EW} + \beta^{p, h}_3 \right. \right. \nonumber \\
& & \hspace{2.5cm} \left. \left. +\beta^{p, h}_{3, EW} \right ] X^{(B a_2, K^*_2)}_h + \left ( \delta_{pu} \alpha^h_2 + \frac{3}{2} \alpha^{p, h}_{3, EW} \right ) X^{(B K^*_2, a_2)}_h \right \},  \\
& & {\cal A}_{B^- \rightarrow a^-_2 K^{*0}_2} =\frac{G_F}{\sqrt{2}} \sum_{p=u,c} \lambda^{(s)}_p \left ( \delta_{pu} \beta^h_2 + \alpha^{p, h}_4 - \frac{1}{2} \alpha^{p, h}_{4, EW} + \beta^{p, h}_3 + \beta^{p, h}_{3, EW} \right ) X^{(B a_2, K^*_2)}_h, \\
& & {\cal A}_{B^0 \rightarrow a^+_2 K^{*-}_2} =\frac{G_F}{\sqrt{2}} \sum_{p=u,c} \lambda^{(s)}_p \left ( \delta_{pu} \alpha^h_1 + \alpha^{p, h}_4, + \alpha^{p, h}_{4, EW} + \beta^{p, h}_3 - \frac{1}{2} \beta^{p, h}_{3, EW} \right ) X^{(B a_2, K^*_2)}_h, \\
& & {\cal A}_{B^0 \rightarrow a^0_2 K^{*0}_2} =\frac{G_F}{2} \sum_{p=u,c} \lambda^{(s)}_p \left [ \left ( - \alpha^{p, h}_4 + \frac{1}{2} \alpha^{p, h}_{4, EW} - \beta^{p, h}_3 + \frac{1}{2} \beta^{p, h}_{3, EW} \right ) X^{(B a_2, K^*_2)}_h \right. \nonumber \\
& & \hspace{2.5cm} \left. + \left ( \delta_{pu} \alpha^h_2 + \frac{3}{2} \alpha^{p, h}_{3, EW} \right ) X^{(B K^*_2, a_2)}_h \right ], \\
& & {\cal A}_{B^- \rightarrow f_2 K^{*-}_2} =\frac{G_F}{\sqrt{2}} \sum_{p=u,c} \lambda^{(s)}_p \left \{ \frac{1}{\sqrt{2}} \left ( \delta_{pu} \alpha^h_2 + 2 \alpha^{p, h}_3 + \frac{1}{2} \alpha^{p, h}_{3, EW} \right ) X^{(B K^*_2, f_2)}_h \right. \nonumber \\
& & \hspace{2.5cm} \left. + \left [ \delta_{pu} \left ( \alpha^h_1 + \beta^h_2 \right ) + \alpha^{p, h}_4 + \alpha^{p, h}_{4, EW} + \beta^{p, h}_3 + \beta^{p,h}_{3, EW} \right ] X^{(B f_2, K^*_2)}_h \right \}, \\
& & {\cal A}_{B^0 \rightarrow f_2 K^{*0}_2} =\frac{G_F}{\sqrt{2}} \sum_{p=u,c} \lambda^{(s)}_p \left [ \frac{1}{\sqrt{2}} \left ( \delta_{pu} \alpha^h_2 + 2 \alpha^{p, h}_3 + \frac{1}{2} \alpha^{p, h}_{3, EW} \right ) X^{(B K^*_2, f_2)}_h \right. \nonumber \\
& & \hspace{2.5cm} \left. + \left ( \alpha^{p, h}_4 - \frac{1}{2} \alpha^{p, h}_{4, EW} + \beta^h_3 - \frac{1}{2} \beta^{p, h}_{3, EW} \right ) X^{(B f_2, K^*_2)}_h \right ], \\
& & {\cal A}_{B^- \rightarrow K^{*-}_2 f^\prime_2} =\frac{G_F}{\sqrt{2}} \sum_{p=u,c} \lambda^{(s)}_p \left ( \delta_{pu} \beta^h_2 + \alpha^{p, h}_3 + \alpha^{p, h}_4 - \frac{1}{2} \alpha^{p, h}_{3, EW} \right. \nonumber \\
& & \hspace{2.5cm} \left. - \frac{1}{2} \alpha^{p, h}_{4, EW} + \beta^{p, h}_3 + \beta^{p, h}_{3, EW} \right ] X^{(B K^*_2, f^\prime_2 )}_h,\\
& & {\cal A}_{B^0 \rightarrow K^{*0}_2 f^\prime_2} =\frac{G_F}{\sqrt{2}} \sum_{p=u,c} \lambda^{(s)}_p \left ( \alpha^{p, h}_3 + \alpha^{p, h}_4 - \frac{1}{2} \alpha^{p, h}_{3, EW} - \frac{1}{2} \alpha^{p, h}_{4, EW} \right. \nonumber \\
& & \hspace{2.5cm} \left. + \beta^{p, h}_3 - \frac{1}{2} \beta^{p, h}_{3, EW} \right ) X^{(B K^*_2, f^\prime)}_h, \\
& & {\cal A}_{B^0 \rightarrow f^\prime_2 f^\prime_2} =\frac{G_F}{\sqrt{2}} \sum_{p=u,c} \lambda^{(d)}_p i f_B f^2_{f^\prime_2} \left ( b^{p,h}_4- \frac{1}{2} b^{p,h}_{4, EW} \right ),
\end{eqnarray}
where the mixing between $f_2(1270)$ and $f^\prime_2(1525)$ have been neglected.

\section{Numerical results}

The decay amplitude for $B \rightarrow TT$ can be decomposed into three components, one for each helicity of the final state, i.e. ${\cal A}_0$, ${\cal A}_+$ and ${\cal A}_-$. The decay rate can be expressed in terms of these amplitudes as
\begin{eqnarray}
\Gamma = \frac{p_c}{8 \pi m^2_B} \left ( |{\cal A}_0|^2 + | {\cal A}_+|^2 + | {\cal A}_-|^2 \right ).
\end{eqnarray}
The branching ratio
\begin{eqnarray}
Br = \frac{\tau_B}{\hbar} \Gamma,
\end{eqnarray}
where $\tau_B$ is the lifetime of $B$ meson. On this basis, we can define the CP averaging branching ratio
\begin{eqnarray}
\frac{1}{2} \left [ Br (B \rightarrow f) + Br (\bar{B} \rightarrow \bar{f} ) \right ],
\end{eqnarray}
and the CP asymmetry
\begin{eqnarray}
A_{CP} = \frac{Br (B \rightarrow f ) - Br (\bar{B} \rightarrow \bar{f} ) }{ Br (B \rightarrow f ) + Br (\bar{B} \rightarrow \bar{f} )},
\end{eqnarray}
where $f$ denotes the final state and $\bar{B} \rightarrow \bar{f}$ is the CP conjugate mode of $B \rightarrow f$. In addition, it is also meaningful to define the longitudinal polarization fraction
\begin{eqnarray}
f_L = \frac{\Gamma_L}{\Gamma},
\end{eqnarray}
where $\Gamma_L = \frac{p_c}{8 \pi m^2_B} |{\cal A}_0 |^2$ is the longitudinal polarization decay rate.

Before performing the actual calculations, we need to summarize all the input parameters involved in this work. Some of the input quantities are collected in TABLE \ref{tab:input}. The decay constants and LCDAs of tensor mesons, and the $B \rightarrow T$ form factors have been specified in Sec.\ref{SecII}, c.f. Eq.(\ref{DCT}), Eqs.(\ref{DAsT1})-(\ref{DAsT8}) and TABLE \ref{tab:FFinputs}, respectively.
As mentioned previously, we shall evaluate the vertex corrections to the decay amplitude at the scale $\mu=m_b$, while the hard spectator and annihilation contributions will be evaluated at the hard collinear scale $\mu_h = \sqrt{\mu \Lambda_h}$ with $\Lambda_h \approx 0.5 {\rm GeV}$. We take the Wilson coefficients $c_i$ at these two scales from Ref.\cite{WilsonC}, detailed in TABLE \ref{tab:WilsonC}. For the effective Wilson coefficients of the magnetic penguin operators in the penguin contractions, we take $c^{eff}_{7\gamma}= 0.299$, $c^{eff}_{8g} = 0.143$\cite{WilsonCadd}.
In addition, the four real parameters $\rho_A$, $\phi_A$, $\rho_H$, $\phi_H$ (introduced in dealing with the endpoint divergence in the annihilation and hard spectator scattering contributions, c.f. Eqs.(\ref{XA}), (\ref{XH})) are involved in our calculations. There are currently no limits on these parameters, so we take $0 \leq \rho_A, \rho_H \leq 1$ and $0 \leq \phi_A, \phi_H \leq 2 \pi$. And for the reduced Plank constant and Fermi coupling constant, we take $\hbar= 6.582 \times 10^{-25} {\rm GeV} \cdot {\rm s}$, $G_F = 1.166 \times 10^{-5} {\rm GeV}^{-2}$\cite{PDG2024}.

With these considerations, we calculate the branching ratios, longitudinal polarization fractions and CP asymmetries for the decays $B \rightarrow a_2 a_2$, $a_2 f_2$, $a_2 K^*_2$, $a_2 f^\prime_2$, $f_2 f_2$, $f_2 K^*_2$, $f_2 f^\prime_2$, $K^*_2 K^*_2$, $K^*_2 f^\prime_2$, $f^\prime_2 f^\prime_2$. The numerical results are collected in TABLE \ref{tab:Btoa2a2}-\ref{tab:Btof2pf2p}, in which $Br_0$ is the longitudinal branching ratio and Avg. denotes an average over the CP-conjugate modes. The uncertainties originate from the variations of decay constants, quark masses, form factors, the $\lambda_B$ parameter for the $B$ meson wave function and the parameters $\rho_A$, $\phi_A$, $\rho_H$, $\phi_H$, added in quadrature. For the branching ratios and CP asymmetries, the total uncertainties are comparable to or even larger than corresponding central values. For the longitudinal polarization fractions, the total uncertainties are roughly $(10-85)\%$. From the perspective of central values, the branching ratios for the decays $B \rightarrow \{a_2 f^\prime_2, f_2 f^\prime_2\}$, $\{a_2f_2, f_2 f_2\}$ and $\{f_2 K^*_2, K^*_2f^\prime_2\}$ are at the order of $10^{-8}$, $10^{-7}$ and $10^{-6}$, respectively. The branching ratios for the decays into $\{a_2 a_2, K^*_2 K^*_2\}$ and $a_2 K^*_2$ are at the order of $10^{-8} - 10^{-7}$ and $10^{-7} - 10^{-6}$, respectively. The branching ratios for the decays into $f^\prime_2 f^\prime_2$ are extremely small, at the order of $10^{-11}$, due to the fact that these modes only receive the contributions from annihilation topologies. The longitudinal polarization fractions are approximately $0.7-1$, implying that all relevant decays are mainly dominated by the longitudinal polarization. Especially, for the $\bar{B}^0/B^0 \rightarrow \{K^{*-}_2 K^{*+}_2, f^\prime_2 f^\prime_2\}$ decays, the longitudinal polarization fractions are equal to $1$, all contributions coming from longitudinal polarization. The CP asymmetry for the $B^\pm \rightarrow a^\pm_2 f_2$ modes is most significant, roughly $-22\%$. The CP asymmetries for the decays $B \rightarrow \{a_2 f^\prime_2, f_2 f^\prime_2, f^\prime_2 f^\prime_2\}$ and $\bar{B}^0/B^0 \rightarrow K^{*-}_2 K^{*+}_2$ are equal to zero.

\begin{table}
\begin{center}
\begin{tabular}{|c|c|c|c|c|}
\hline
\multicolumn{5}{|c|}{$B$ mesons\cite{PDG2024,LB}} \\
\hline
$B$ & $m_B({\rm GeV}$) & $\tau_B({\rm s})$ &
$f_B({\rm GeV})$ & $\lambda_B({\rm GeV})$ \\
\hline
$B^{\pm}$ & $5.279$ & $1.638 \times 10^{-12}$ &
$0.1900\pm 0.0013$ & $0.383\pm 0.153$ \\
\hline
$B^0$ & $5.280$ & $1.519 \times 10^{-12}$ &
$0.1900\pm 0.0013$ & $0.383\pm 0.153$ \\
\hline
\multicolumn{5}{|c|}{The masses of tensor mesons\cite{PDG2024}}  \\
\hline
 $m_{a_2}({\rm GeV})$ & $m_{f_2}({\rm GeV})$  & $m_{K^{*\pm}_2}({\rm GeV})$ & $m_{K^{*0}_2}({\rm GeV})$ &
$m_{f^\prime_2}({\rm GeV})$   \\
\hline
$1.318$ & $1.275$ & $1.427$ & $1.432$ & $1.517$ \\
\hline
\multicolumn{5}{|c|}{The masses of quarks\cite{PDG2024}} \\
\hline
$m_u({\rm GeV})$ & $m_d$({\rm GeV}) & $m_s$ ({\rm GeV})& $m_c$({\rm GeV}) & $m_b$({\rm GeV})  \\
\hline
$0.00216^{+0.00049}_{-0.00026}$ & $0.00467^{+0.00048}_{-0.00017}$ & $0.0934^{+0.0086}_{-0.0034}$ & $1.27 \pm 0.02$ & $4.18^{+0.03}_{-0.02}$  \\
\hline
\multicolumn{5}{|c|}{Wolfenstein parameters\cite{PDG2024}} \\
\hline
$A$ & $\lambda$ & $\bar{\rho}$ & $\bar{\eta}$ & ---\\
\hline
$0.826$ & $0.225$ & $0.159$ & $0.352$ & --- \\
\hline
\end{tabular}
\end{center}
\caption{
Input parameters relevant to the $B$ mesons, the masses of tensor mesons and quarks, and the Wolfenstein parameters.
} \label{tab:input}
\end{table}

\begin{table}
\centering
\begin{tabular}{|c|c|c|c|c|c|c|c|c|c|c|}
\hline Scale & $c_1$ & $c_2$ & $c_3$ & $c_4$ & $c_5$ & $c_6$ & $c_7/\alpha$ & $c_8/\alpha$ & $c_9/\alpha$ & $c_{10}/\alpha$  \\
\hline
$\mu$ & $1.0813$ & $-0.1903$ & $0.0137$ & $-0.0357$ & $0.0087$ & $-0.0419$ & $-0.0026$ & $0.0618$ & $-1.2423$ & $0.2283$ \\
\hline
$\mu_h$ & $1.1820$ & $-0.3700$ & $0.0274$ & $-0.0618$ & $0.0105$ & $-0.0854$ & $-0.0147$ & $0.1317$ & $-1.3526$ & $0.4015$ \\
\hline
\end{tabular}
\caption{Next to leading order Wilson coefficients $c_i$ in the NDR scheme at the scale $\mu$ and $\mu_h$, used as inputs\cite{WilsonC}.}\label{tab:WilsonC}
\end{table}

\begin{table}
\centering
\begin{tabular}{|c|c|c|c|c|}
\hline
Decays & $Br_0(10^{-7})$ & $Br(10^{-7})$ & $f_L$ & $A_{CP}(\%)$ \\
\hline
$B^- \rightarrow a^-_2 a^0_2$ & $1.22^{+1.26}_{-1.06}$ & $1.24^{+1.28}_{-0.98}$ & $0.987^{+0.013}_{-0.212}$ & \\
\cline{1-4}
$B^+ \rightarrow a^+_2 a^0_2$ & $1.23^{+1.26}_{-1.06}$ & $1.25^{+1.28}_{-0.98}$ & $0.987^{+0.013}_{-0.209}$ & $-0.41^{+0.39}_{-1.03}$ \\
\cline{1-4}
Avg. & $1.23^{+1.26}_{-1.06}$ & $1.24^{+1.28}_{-0.98}$ & $0.987^{+0.013}_{-0.210}$ &  \\
\hline
\hline
$\bar{B}^0 \rightarrow a^-_2 a^+_2$ & $0.70^{+5.79}_{-0.70}$ & $0.85^{+8.22}_{-0.85}$ & $0.824^{+0.176}_{-0.577}$ & \\
\cline{1-4}
$B^0 \rightarrow a^+_2 a^-_2$ & $0.78^{+5.40}_{-0.78}$ & $0.97^{+7.88}_{-0.97}$ & $0.811^{+0.189}_{-0.690}$ & $-6.56^{+72.96}_{-93.44}$ \\
\cline{1-4}
Avg. & $0.74^{+5.52}_{-0.57}$ & $0.91^{+8.00}_{-0.69}$ & $0.818^{+0.182}_{-0.367}$ &  \\
\hline
\hline
$\bar{B}^0 \rightarrow a^0_2 a^0_2$ & $1.17^{+2.28}_{-1.17}$ & $1.18^{+2.74}_{-1.18}$ & $0.989^{+0.011}_{-0.236}$ & \\
\cline{1-4}
$B^0 \rightarrow a^0_2 a^0_2$ & $0.89^{+1.79}_{-0.89}$ & $0.97^{+2.13}_{-0.94}$ & $0.910^{+0.090}_{-0.537}$ & $9.58^{+58.22}_{-79.38}$ \\
\cline{1-4}
Avg. & $1.03^{+1.74}_{-0.87}$ & $1.08^{+2.24}_{-0.80}$ & $0.949^{+0.051}_{-0.357}$ &  \\
\hline
\end{tabular}
\caption{The branching ratios, longitudinal polarizaiton fractions and CP asymmetries for $B \rightarrow a_2 a_2$ decays }\label{tab:Btoa2a2}
\end{table}

\begin{table}
\centering
\begin{tabular}{|c|c|c|c|c|}
\hline
Decays & $Br_0(10^{-7})$ & $Br(10^{-7})$ & $f_L$ & $A_{CP}(\%)$ \\
\hline
$B^- \rightarrow a^-_2 f_2$ & $3.75^{+18.84}_{-3.75}$ & $4.71^{+27.93}_{-4.71}$ & $0.795^{+0.205}_{-0.500}$ & \\
\cline{1-4}
$B^+ \rightarrow a^+_2 f_2$ & $6.78^{+26.11}_{-6.78}$ & $7.30^{+34.21}_{-7.30}$ & $0.928^{+0.072}_{-0.490}$ & $-21.58^{+85.42}_{-37.06}$ \\
\cline{1-4}
Avg. & $5.26^{+22.39}_{-5.26}$ & $6.01^{+31.03}_{-6.01}$ & $0.862^{+0.138}_{-0.300}$ &  \\
\hline
\hline
$\bar{B}^0 \rightarrow a^0_2 f_2$ & $1.90^{+10.43}_{-1.90}$ & $2.26^{+14.27}_{-2.26}$ & $0.842^{+0.158}_{-0.580}$ & \\
\cline{1-4}
$B^0 \rightarrow a^0_2 f_2$ & $1.89^{+9.58}_{-1.89}$ & $2.29^{+13.67}_{-2.29}$ & $0.827^{+0.123}_{-0.516}$ & $-0.67^{+70.50}_{-67.05}$ \\
\cline{1-4}
Avg. & $1.90^{+9.99}_{-1.90}$ & $2.27^{+13.96}_{-2.27}$ & $0.834^{+0.104}_{-0.513}$ &  \\
\hline
\end{tabular}
\caption{The branching ratios, longitudinal polarizaiton fractions and CP asymmetries for $B \rightarrow a_2 f_2$ decays }\label{tab:Btoa2f2}
\end{table}

\begin{table}
\centering
\begin{tabular}{|c|c|c|c|c|}
\hline
Decays & $Br_0(10^{-7})$ & $Br(10^{-7})$ & $f_L$ & $A_{CP}(\%)$ \\
\hline
$B^- \rightarrow a^0_2 K^{*-}_2$ & $5.64^{+59.79}_{-5.64}$ & $7.33^{+87.72}_{-7.33}$ & $0.770^{+0.193}_{-0.301}$ & \\
\cline{1-4}
$B^+ \rightarrow a^0_2 K^{*+}_2$ & $4.46^{+54.01}_{-4.46}$ & $6.45^{+82.85}_{-6.45}$ & $0.692^{+0.308}_{-0.457}$ & $6.41^{+39.54}_{-87.90}$ \\
\cline{1-4}
Avg. & $5.05^{+56.83}_{-5.05}$ & $6.89^{+85.24}_{-6.89}$ & $0.731^{+0.246}_{-0.297}$ &  \\
\hline
\hline
$B^- \rightarrow a^-_2 \bar{K}^{*0}_2$ & $9.95^{+115.00}_{-9.95}$ & $13.75^{+172.40}_{-13.75}$ & $0.724^{+0.267}_{-0.268}$ & \\
\cline{1-4}
$B^+ \rightarrow a^+_2 K^{*0}_2$ & $9.72^{+114.37}_{-9.72}$ & $13.54^{+171.85}_{-13.54}$ & $0.718^{+0.279}_{-0.271}$ & $0.77^{+1.35}_{-3.69}$ \\
\cline{1-4}
Avg. & $9.84^{+114.68}_{-9.84}$ & $13.65^{+172.12}_{-13.65}$ & $0.721^{+0.273}_{-0.269}$ &  \\
\hline
\hline
$\bar{B}^0 \rightarrow a^+_2 K^{*-}_2$ & $12.03^{+111.74}_{-12.03}$ & $15.65^{+164.85}_{-15.65}$ & $0.769^{+0.231}_{-0.429}$ & \\
\cline{1-4}
$B^0 \rightarrow a^-_2 K^{*+}_2$ & $11.83^{+112.20}_{-11.83}$ & $15.41^{+165.21}_{-15.41}$ & $0.768^{+0.227}_{-0.408}$ & $0.76^{+14.21}_{-6.64}$ \\
\cline{1-4}
Avg. & $11.93^{+111.97}_{-11.93}$ & $15.53^{+165.03}_{-15.53}$ & $0.768^{+0.231}_{-0.418}$ &  \\
\hline
\hline
$\bar{B}^0 \rightarrow a^0_2 \bar{K}^{*0}_2$ & $5.28^{+53.69}_{-5.28}$ & $7.18^{+80.62}_{-7.18}$ & $0.735^{+0.265}_{-0.354}$ & \\
\cline{1-4}
$B^0 \rightarrow a^0_2 K^{*0}_2$ & $5.66^{+56.07}_{-5.66}$ & $7.46^{+82.66}_{-7.46}$ & $0.759^{+0.216}_{-0.348}$ & $-1.90^{+45.53}_{-21.86}$ \\
\cline{1-4}
Avg. & $5.47^{+54.87}_{-5.47}$ & $7.32^{+81.63}_{-7.32}$ & $0.747^{+0.247}_{-0.336}$ &  \\
\hline
\end{tabular}
\caption{The branching ratios, longitudinal polarizaiton fractions and CP asymmetries for $B \rightarrow a_2 K^*_2$ decays }\label{tab:Btoa2K2}
\end{table}

\begin{table}
\centering
\begin{tabular}{|c|c|c|c|c|}
\hline
Decays & $Br_0(10^{-8})$ & $Br(10^{-8})$ & $f_L$ & $A_{CP}(\%)$ \\
\hline
$B^- \rightarrow a^-_2 f^\prime_2$ & $3.85^{+5.73}_{-3.85}$ & $4.05^{+6.18}_{-4.05}$ & $0.950^{+0.043}_{-0.075}$ & \\
\cline{1-4}
$B^+ \rightarrow a^+_2 f^\prime_2$ & $3.85^{+5.73}_{-3.85}$ & $4.05^{+6.18}_{-4.05}$ & $0.950^{+0.043}_{-0.075}$ & $0$ \\
\cline{1-4}
Avg. & $3.85^{+5.73}_{-3.85}$ & $4.05^{+6.18}_{-4.05}$ & $0.950^{+0.043}_{-0.075}$ &  \\
\hline
\hline
$\bar{B}^0 \rightarrow a^0_2 f^\prime_2$ & $1.79^{+2.66}_{-1.79}$ & $1.88^{+2.87}_{-1.88}$ & $0.950^{+0.043}_{-0.075}$ & \\
\cline{1-4}
$B^0 \rightarrow a^0_2 f^\prime_2$ & $1.79^{+2.66}_{-1.79}$ & $1.88^{+2.87}_{-1.88}$ & $0.950^{+0.043}_{-0.075}$ & $0$ \\
\cline{1-4}
Avg. & $1.79^{+2.66}_{-1.79}$ & $1.88^{+2.87}_{-1.88}$ & $0.950^{+0.043}_{-0.075}$ &  \\
\hline
\end{tabular}
\caption{The branching ratios, longitudinal polarizaiton fractions and CP asymmetries for $B \rightarrow a_2 f^\prime_2$ decays }\label{tab:Btoa2f2p}
\end{table}

\begin{table}
\centering
\begin{tabular}{|c|c|c|c|c|}
\hline
Decays & $Br_0(10^{-7})$ & $Br(10^{-7})$ & $f_L$ & $A_{CP}(\%)$ \\
\hline
$\bar{B}^0 \rightarrow f_2 f_2$ & $2.94^{+4.71}_{-2.94}$ & $3.29^{+6.12}_{-3.29}$ & $0.895^{+0.105}_{-0.378}$ & \\
\cline{1-4}
$B^0 \rightarrow f_2 f_2$ & $4.38^{+7.21}_{-4.38}$ & $4.49^{+8.45}_{-4.49}$ & $0.977^{+0.023}_{-0.112}$ & $-15.40^{+57.60}_{-42.16}$ \\
\cline{1-4}
Avg. & $3.66^{+5.78}_{-3.66}$ & $3.89^{+7.19}_{-3.78}$ & $0.936^{+0.064}_{-0.221}$ &  \\
\hline
\end{tabular}
\caption{The branching ratios, longitudinal polarizaiton fractions and CP asymmetries for $B \rightarrow f_2 f_2$ decays }\label{tab:Btof2f2}
\end{table}

\begin{table}
\centering
\begin{tabular}{|c|c|c|c|c|}
\hline
Decays & $Br_0(10^{-6})$ & $Br(10^{-6})$ & $f_L$ & $A_{CP}(\%)$ \\
\hline
$B^- \rightarrow f_2 K^{*-}_2$ & $4.56^{+17.89}_{-4.56}$ & $5.15^{+24.59}_{-5.15}$ & $0.885^{+0.115}_{-0.571}$ & \\
\cline{1-4}
$B^+ \rightarrow f_2 K^{*+}_2$ & $4.38^{+17.53}_{-4.38}$ & $4.99^{+24.27}_{-4.99}$ & $0.877^{+0.123}_{-0.578}$ & $1.57^{+2.15}_{-5.24}$ \\
\cline{1-4}
Avg. & $4.47^{+17.71}_{-4.47}$ & $5.07^{+24.43}_{-5.07}$ & $0.881^{+0.119}_{-0.573}$ &  \\
\hline
\hline
$\bar{B}^0 \rightarrow f_2 \bar{K}^{*0}_2$ & $3.65^{+15.90}_{-3.65}$ & $4.14^{+21.97}_{-4.14}$ & $0.880^{+0.120}_{-0.711}$ & \\
\cline{1-4}
$B^0 \rightarrow f_2 K^{*0}_2$ & $3.49^{+15.46}_{-3.49}$ & $4.01^{+21.60}_{-4.01}$ & $0.870^{+0.130}_{-0.758}$ & $1.62^{+7.21}_{-11.41}$ \\
\cline{1-4}
Avg. & $3.57^{+15.68}_{-3.57}$ & $4.08^{+21.79}_{-4.08}$ & $0.875^{+0.125}_{-0.715}$ &  \\
\hline
\end{tabular}
\caption{The branching ratios, longitudinal polarizaiton fractions and CP asymmetries for $B \rightarrow f_2 K^*_2$ decays }\label{tab:Btof2K2}
\end{table}

\begin{table}
\centering
\begin{tabular}{|c|c|c|c|c|}
\hline
Decays & $Br_0(10^{-8})$ & $Br(10^{-8})$ & $f_L$ & $A_{CP}(\%)$ \\
\hline
$\bar{B}^0 \rightarrow f_2 f^\prime_2$ & $3.76^{+5.75}_{-3.76}$ & $4.00^{+6.27}_{-4.00}$ & $0.941^{+0.050}_{-0.100}$ & \\
\cline{1-4}
$B^0 \rightarrow f_2 f^\prime_2$ & $3.76^{+5.75}_{-3.76}$ & $4.00^{+6.27}_{-4.00}$ & $0.941^{+0.050}_{-0.100}$ & $0$ \\
\cline{1-4}
Avg. & $3.76^{+5.75}_{-3.76}$ & $4.00^{+6.27}_{-4.00}$ & $0.941^{+0.050}_{-0.100}$ &  \\
\hline
\end{tabular}
\caption{The branching ratios, longitudinal polarizaiton fractions and CP asymmetries for $B \rightarrow f_2 f^\prime_2$ decays }\label{tab:Btof2f2p}
\end{table}

\begin{table}
\centering
\begin{tabular}{|c|c|c|c|c|}
\hline
Decays & $Br_0(10^{-7})$ & $Br(10^{-7})$ & $f_L$ & $A_{CP}(\%)$ \\
\hline
$B^- \rightarrow K^{*-}_2 K^{*0}_2$ & $0.27^{+3.95}_{-0.27}$ & $0.42^{+5.98}_{-0.42}$ & $0.655^{+0.345}_{-0.347}$ & \\
\cline{1-4}
$B^+ \rightarrow K^{*+}_2 \bar{K}^{*0}_2$ & $0.50^{+4.51}_{-0.50}$ & $0.62^{+6.50}_{-0.62}$ & $0.796^{+0.183}_{-0.548}$ & $-19.49^{+78.47}_{-27.60}$ \\
\cline{1-4}
Avg. & $0.38^{+4.22}_{-0.38}$ & $0.52^{+6.24}_{-0.52}$ & $0.725^{+0.275}_{-0.274}$ &  \\
\hline
\hline
$\bar{B}^0 \rightarrow K^{*-}_2 K^{*+}_2$ & $3.57^{+58.97}_{-3.57}$ & $3.57^{+58.97}_{-3.57}$ & $1$ & \\
\cline{1-4}
$B^0 \rightarrow K^{*+}_2 K^{*-}_2$ & $3.57^{+58.97}_{-3.57}$ & $3.57^{+58.97}_{-3.57}$ & $1$ & $0$ \\
\cline{1-4}
Avg. & $3.57^{+58.97}_{-3.57}$ & $3.57^{+58.97}_{-3.57}$ & $1$ &  \\
\hline
\hline
$\bar{B}^0 \rightarrow \bar{K}^{*0}_2 K^{*0}_2$ & $1.99^{+68.75}_{-1.99}$ & $2.13^{+68.93}_{-2.13}$ & $0.937^{+0.063}_{-0.324}$ & \\
\cline{1-4}
$B^0 \rightarrow K^{*0}_2 \bar{K}^{*0}_2$ & $2.03^{+69.02}_{-2.03}$ & $2.14^{+69.20}_{-2.14}$ & $0.945^{+0.055}_{-0.200}$ & $-0.40^{+7.49}_{-25.55}$ \\
\cline{1-4}
Avg. & $2.01^{+68.74}_{-2.01}$ & $2.13^{+68.91}_{-2.13}$ & $0.941^{+0.059}_{-0.219}$ &  \\
\hline
\end{tabular}
\caption{The branching ratios, longitudinal polarizaiton fractions and CP asymmetries for $B \rightarrow K^*_2 K^*_2$ decays }\label{tab:BtoK2K2}
\end{table}

\begin{table}
\centering
\begin{tabular}{|c|c|c|c|c|}
\hline
Decays & $Br_0(10^{-6})$ & $Br(10^{-6})$ & $f_L$ & $A_{CP}(\%)$ \\
\hline
$B^- \rightarrow K^{*-}_2 f^\prime_2$ & $3.14^{+13.05}_{-3.14}$ & $3.63^{+18.00}_{-3.63}$ & $0.866^{+0.134}_{-0.614}$ & \\
\cline{1-4}
$B^+ \rightarrow K^{*+}_2 f^\prime_2$ & $3.09^{+12.99}_{-3.09}$ & $3.58^{+17.94}_{-3.58}$ & $0.864^{+0.136}_{-0.598}$ & $0.61^{+1.91}_{-1.09}$ \\
\cline{1-4}
Avg. & $3.12^{+13.02}_{-3.12}$ & $3.60^{+17.97}_{-3.60}$ & $0.865^{+0.135}_{-0.606}$ &  \\
\hline
\hline
$\bar{B}^0 \rightarrow K^{*0}_2 f^\prime_2$ & $2.87^{+11.97}_{-2.87}$ & $3.32^{+16.55}_{-3.32}$ & $0.865^{+0.135}_{-0.601}$ & \\
\cline{1-4}
$B^0 \rightarrow \bar{K}^{*0}_2 f^\prime_2$ & $2.83^{+11.92}_{-2.83}$ & $3.29^{+16.51}_{-3.29}$ & $0.863^{+0.137}_{-0.610}$ & $0.52^{+1.12}_{-0.53}$ \\
\cline{1-4}
Avg. & $2.85^{+11.95}_{-2.85}$ & $3.30^{+16.53}_{-3.30}$ & $0.864^{+0.136}_{-0.605}$ &  \\
\hline
\end{tabular}
\caption{The branching ratios, longitudinal polarizaiton fractions and CP asymmetries for $B \rightarrow K^*_2 f^\prime_2$ decays }\label{tab:BtoK2f2p}
\end{table}

\begin{table}
\centering
\begin{tabular}{|c|c|c|c|c|}
\hline
Decays & $Br_0(10^{-11})$ & $Br(10^{-11})$ & $f_L$ & $A_{CP}(\%)$ \\
\hline
$\bar{B}^0 \rightarrow f^\prime_2 f^\prime_2$ & $5.43^{+60.78}_{-5.43}$ & $5.43^{+60.78}_{-5.43}$ & $1$ & \\
\cline{1-4}
$B^0 \rightarrow f^\prime_2 f_2$ & $5.43^{+60.78}_{-5.43}$ & $5.43^{+60.78}_{-5.43}$ & $1$ & $0$ \\
\cline{1-4}
Avg. & $5.43^{+60.78}_{-5.43}$ & $5.43^{+60.78}_{-5.43}$ & $1$ &  \\
\hline
\end{tabular}
\caption{The branching ratios, longitudinal polarizaiton fractions and CP asymmetries for $B \rightarrow f^\prime_2 f^\prime_2$ decays }\label{tab:Btof2pf2p}
\end{table}

\section{Summary}

In this study, we investigate the $B \rightarrow TT$ decays in the QCD factorization approach, with $T$ denoting the tensor mesons $a_2(1320)$, $K^*_2(1430)$, $f_2(1270)$ and $f^\prime_2(1525)$. After a brief review of the physical properties of tensor mesons used in our calculations such as the decay constants, form factors and LCDAs, we work out the next to leading order corrections to these decays. In specific, the nonfactorizable contributions, including the vertex corrections, penguin contractions, hard spectator scattering and annihilation topologies, are directly obtained from the $B \rightarrow TV$ decays by making some suitable replacements, due to the fact that the longitudinal and transverse helicity projection operators for the tensor meson are very similar to the projectors for the vector meson. Then, the branching ratios, longitudinal polarization fractions, and CP asymmetries for all relevant channels are predicted systematically. The uncertainties originate from the variations of decay constants, quark masses, form factors, the $\lambda_B$ parameter for the $B$ meson wave function and the parameters $\rho_A$, $\phi_A$, $\rho_H$, $\phi_H$. Added in quadrature, the total uncertainties are comparable to or even larger than corresponding central values for the branching ratios and CP asymmetries.

From the perspective of central values, the branching ratios for the decays $B \rightarrow \{a_2 f^\prime_2, f_2 f^\prime_2\}$, $\{a_2f_2, f_2 f_2\}$ and $\{f_2 K^*_2, K^*_2f^\prime_2\}$ are at the order of $10^{-8}$, $10^{-7}$ and $10^{-6}$, respectively. The branching ratios for the decays into $\{a_2 a_2, K^*_2 K^*_2\}$ and $a_2 K^*_2$ are at the order of $10^{-8} - 10^{-7}$ and $10^{-7} - 10^{-6}$, respectively. The branching ratios for the decays into $f^\prime_2 f^\prime_2$ are extremely small, at the order of $10^{-11}$, due to the fact that these modes only receive the contributions from annihilation topologies. The longitudinal polarization fractions are approximately $0.7-1$, implying that all relevant decays are mainly dominated by the longitudinal polarization. Particularly, for the $\bar{B}^0/B^0 \rightarrow \{K^{*-}_2 K^{*+}_2, f^\prime_2 f^\prime_2\}$ decays, the longitudinal polarization fractions are equal to $1$, all contributions coming from longitudinal polarization. The CP asymmetry for the $B^\pm \rightarrow a^\pm_2 f_2$ modes is most significant, roughly $-22\%$. The CP asymmetries for the decays $B \rightarrow \{a_2 f^\prime_2, f_2 f^\prime_2, f^\prime_2 f^\prime_2\}$ and $\bar{B}^0/B^0 \rightarrow K^{*-}_2 K^{*+}_2$ are equal to zero. These results may be tested by more precise experiments in the future.

%\begin{acknowledgments}

%This work was supported  in part by the National Natural Science
%Foundation of China under Grant No. 12147214.

%\end{acknowledgments}

\end{document}